\pdfoutput=1 
\documentclass[useAMS,usenatbib,letters]{mnras}

\usepackage{amsmath}
\usepackage{amssymb}
\usepackage{times,txfonts}
\usepackage{graphicx}

\usepackage{newtxtext,newtxmath}
\usepackage[T1]{fontenc}
\usepackage{ae,aecompl}
\usepackage{float}

\hypersetup{draft}

\newcommand{\ud}{\ensuremath{\mathrm{d}}}
\newcommand{\pd}{\partial}

\title[Neutrino-Driven Supernova Aided by Magnetic Fields]{A 3D Simulation of a Neutrino-Driven Supernova Explosion Aided By Convection and Magnetic Fields}

\author[M\"uller \& Varma]{
Bernhard M\"uller$^{1,2}$\thanks{E-mail: bernhard.mueller@monash.edu}
and Vishnu Varma$^{1}$
\\
$^{1}$School of Physics and Astronomy, Monash University, VIC 3800, Australia\\
$^{2}$ ARC Centre of Excellence for Gravitational Wave Discovery -- OzGrav
}

\begin{document}

\label{firstpage}
\pagerange{\pageref{firstpage}--\pageref{lastpage}}

\maketitle
             
\begin{abstract}
We study the impact of a small-scale dynamo in core-collapse
supernovae using a 3D neutrino magnetohydrodynamics simulation of a
$15 M_\odot$ progenitor.  The weak seed field is amplified
exponentially in the gain region once neutrino-driven convection
develops, and remains dominated by small-scale structures. About
$250\, \mathrm{ms}$ after bounce, the field energy in the gain region
reaches $\mathord{\sim} 50\%$ of kinetic equipartition. This supports
the development of a neutrino-driven explosion with modest global
anisotropy, which does not occur in a corresponding model without
magnetic fields. Our results suggest that magnetic fields may play a
beneficial subsidiary role in neutrino-driven supernovae even without
rapid progenitor rotation. Further investigation into the nature of
magnetohydrodynamic turbulence in the supernova core is required.
\end{abstract}


\section{Introduction}
\label{sec:intro}
Magnetic field effects pervade many astrophysical
fluid dynamics problems such as stellar surface
convection, stellar winds, and star formation. It has
long been speculated that magnetic fields also play
a critical role in some core-collapse supernova explosions
of massive stars. The idea of tapping
the rotational energy stored in the supernova 
core using strong magnetic fields has a long history 
\citep[e.g.,][]{meier_76,bisnovatyi_76,akiyama_03,dessart_07a}.
In recent years models of such magnetorotational
explosions have matured considerably, and three-dimensional (3D)
simulations based on rapidly rotating stellar progenitor 
models are now available 
\citep{winteler_12,moesta_14b,obergaulinger_17,kuroda_20}.
Since strong magnetic fields in the progenitor star would 
lead to effective core spin-down, magnetorotational
explosions require some amplification mechanism to generate
strong large-scale magnetic fields on short time
scales after core collapse, such as the
magnetorotational instability \citep{balbus_91,akiyama_03}
or an $\alpha$-$\Omega$ dynamo  in the
proto-neutron star (PNS) convection zone \citep{raynaud_20}.
Despite progress in understanding these amplification
processes by means of idealized local and global simulations
and analytic theory
\citep{obergaulinger_09,masada_12,sawai_13,guilet_15,moesta_15,raynaud_20,masada_20}, the stellar pre-collapse rotation
rate remains a major unknown for this supernova mechanism.
Current stellar evolution models including magnetic torques
\citep{heger_05}
predict core spin rates that are too low for 
magnetorotational explosions, and still underestimate
core spin-down in the case of low-mass
red giants \citep{cantiello_14} for which asteroseismic
measurements are available. Thus, magnetorotational 
explosions are likely rare and probably only explain
``hypernovae'' with unusually high explosion energies.
For the majority of massive stars with moderate or
slow core rotation, the neutrino-driven
mechanism \citep{mueller_16b} remains the favored
scenario.

\begin{figure*}
     \centering
     \includegraphics[width=\linewidth]{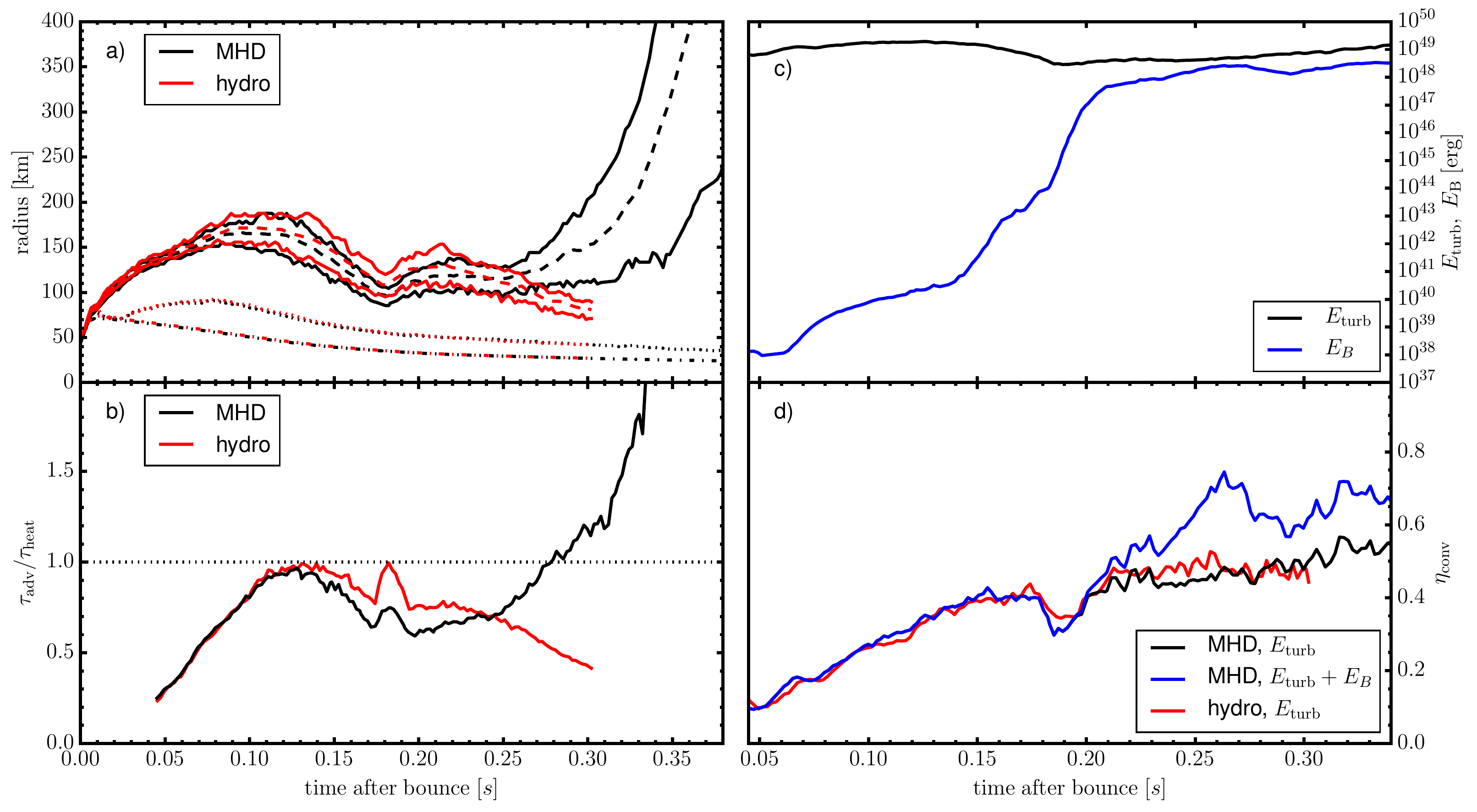}
     \caption{Evolution of the
     MHD model (black/blue curves) and the hydro model
     (red). a) Maximum, minimum (solid)
     and average shock radius (dashed), gain radius
     (dotted), and proto-neutron star radius
     (dash-dotted). b) Critical time
     scale ratio $\tau_\mathrm{adv}/\tau_\mathrm{heat}$.
     c) Turbulent kinetic energy
     $E_\mathrm{turb}$ (black) and magnetic
     energy in the gain region (blue) in
     the MHD model. d) Efficiency
     $\eta_\mathrm{conv}$ for the conversion
     of neutrino heating into turbulent energy kinetic
     energy (black/red)  or total turbulent energy
     including magnetic fields (blue).}
     \label{fig:evol}
 \end{figure*}

Magnetohydrodynamic (MHD) effects in non-rotating
or slowly-rotating progenitors have received less attention,
though a few studies have explored the amplification
and dynamical role of Alfv\'en waves \citep{suzuki_08,guilet_11} and field
amplification by the standing accretion shock instability
\citep{endeve_10,endeve_12}. Supernova simulations 
of non-rotating progenitors with MHD and neutrino transport 
have so far been conducted in axisymmetry (2D) 
only \citep{obergaulinger_14}. These models have indicated that
for strong fields of $\mathord{\sim}10^{12}\, \mathrm{G}$, MHD effects could play an 
auxiliary role in  neutrino-driven explosions by 
facilitating the formation of large high-entropy bubbles.

However, MHD effects could play a more important and more generic role
in neutrino-driven supernovae than these 2D simulations suggested,
since dynamo field amplification cannot operate in 2D
\citep{cowling_33}. More efficient
field amplification might occur in 3D by a small-scale turbulent
dynamo. A small-scale dynamo was in fact seen in idealized 3D
simulations of the standing accretion shock instability
by \citet{endeve_12}, though the fields did not become dynamically
significant in their study. Moreover, conventional estimates for the
field strengths in the cores and inner shells of massive stars could
be too pessimistic. The magnetic field strengths of
$10^3\texttt{-}10^9\, \mathrm{G}$ in white dwarfs \citep{ferrario_15}
may not be indicative of the conditions in massive stars at the
pre-collapse stage, where convective burning could generate strong
small-scale fields via a turbulent dynamo.  Considering ubiquitous
observations of magnetic field strengths close to kinetic
equipartition in similar settings \citep{christensen_09,brun_17}, one
should expect fields of order $10^{10\texttt{-}11}\,\mathrm{G}$ in the
innermost active burning shells at collapse. Here we explore the
amplification of such seed fields by a small-scale dynamo and their
interplay with neutrino heating and the hydrodynamic instabilities in
a progenitor with a moderate rotation rate for the first time in a 3D
MHD simulation with neutrino transport.

\section{Progenitor Model and Initial Conditions}
We simulate the collapse of the $15 M_\odot$
model m15b6 from \citet{heger_05}, whose evolution up to
collapse  has been calculated assuming magnetic torques.
The progenitor has a central rotation rate of 
$0.05\, \mathrm{rad}\, \mathrm{s}^{-1}$, which translates into a neutron star birth spin period
of $11 \, \mathrm{ms}$ assuming that the collapsing
core does not exchange angular momentum with the ejecta
during the explosion. The neutron star's rotational energy of $\mathord{\sim} 2 \times 10^{50}\, \mathrm{erg}$
would thus be too small to power a supernova
 with normal energy by MHD effects alone.

We perform two simulations with and without magnetic
fields.  Following \citet{obergaulinger_17}, we assume a 
dipolar field geometry given by the
vector potential,
\begin{equation}
    (A_r,A_\theta,A_\varphi)=
    (r B_{\mathrm{t},0}(r) \cos \theta,0,
    r/2\times B_{\mathrm{p},0}(r) \sin \theta),
\end{equation}
in terms of the radius-dependent poloidal and toroidal field strengths $B_{\mathrm{p},0}$ and $B_{\mathrm{t},0}$.
Realistic seed fields are likely dominated by smaller scales, 
but in default of better pre-collapse models, the assumption
of a dipolar geometry appears justified as our findings 
do not appear to hinge on the large-scale structure of the field.
In order not to overestimate the  impact
of magnetic fields, we reduce the 
poloidal and toroidal field strengths
$B_{\mathrm{p},\mathrm{prog}}$ and $B_{\mathrm{t},\mathrm{prog}}$
in the progenitor by a factor of $10^4$, i.e.,
$B_{\mathrm{p},0}=10^{-4} B_{\mathrm{p},\mathrm{prog}}$ and
 $B_{\mathrm{t},0}=10^{-4} B_{\mathrm{t},\mathrm{prog}}$.
In the progenitor,
$B_{\mathrm{p},\mathrm{prog}}$ and $B_{\mathrm{t},\mathrm{prog}}$
reach values of $5\times10^9\, \mathrm{G}$ and $10^6\,\mathrm{G}$ in non-convective
regions as predicted by the Tayler-Spruit
dynamo \citep{spruit_02}. Inside convective regions
one expects values of
$B_{\mathrm{p},\mathrm{prog}}$ and $B_{\mathrm{t},\mathrm{prog}}$
close to
kinetic equipartition, which translates into
a plasma beta (defined as the ratio
of thermal to magnetic pressure) of $\beta=10^4$ for the
typical convective Mach numbers of $\mathord{\sim}10^{-2}$ in the innermost burning shells at collapse \citep{collins_18}. 
This would imply rather strong
fields of up to $3\times 10^{12}\,\mathrm{G}$ inside
a small central region of radius $40 \, \mathrm{km}$
and $6\times 10^{10}\,\mathrm{G}$ in the oxygen 
shell, but after rescaling by a factor  $10^{-4}$,
the seed fields can clearly not play any dynamical 
role after collapse without dynamo field
amplification.

\section{Numerical Methods}
The simulations have been conducted
with the Newtonian version of the 
\textsc{CoCoNuT-FMT} code \citep{mueller_15a}
using an effective gravitational potential
(Case~Arot) from \citet{mueller_08}.
The code has been updated to solve
the Newtonian MHD equations 
using the extended HLLC solver of
\citet{gurski_04,miyoshi_05}
and an energy-conserving variant of hyperbolic divergence
cleaning \citep{dedner_02} following ideas from \citet{tricco_16}. Excluding
terms for neutrino interactions and nuclear
reactions, the extended system of MHD equations
for the density $\rho$, velocity $\mathbf{v}$,
magnetic field $\mathbf{B}$,  total
energy density $e$, and the Lagrange
multiplier $\hat{\psi}$ reads,
\begin{align}
\pd_t \rho
+\nabla \cdot \rho \mathbf{v}
&=
0,
\\
\pd_t (\rho \mathbf v)
+\nabla \cdot \left(\rho \mathbf{v}\mathbf{v}-
\frac{\mathbf{B} \mathbf{B}}{4\pi}
+P_\mathrm{t}\mathcal{I}
\right)
&=
\rho \mathbf{g}
-
\frac{(\nabla \cdot\mathbf{B}) \mathbf{B}}{4\pi}
,
\\
\pd_t e+
\nabla \cdot 
\left[(e+P_\mathrm{t})\mathbf{v}
-\frac{\mathbf{B} (\mathbf{v}\cdot\mathbf{B})}{4\pi}
\right]
&=
\rho \mathbf{g}\cdot \mathbf{v}
,
\\
\pd_t \mathbf{B} +\nabla \cdot (\mathbf{v}\mathbf{B}-\mathbf{B}\mathbf{v})
+\nabla \cdot (c_\mathrm{h} \hat{\psi})
&=0
\\
\pd_t \hat{\psi}
+c_\mathrm{h} \nabla \cdot \mathbf{B}
&=-\hat{\psi}/\tau.
\end{align}
where $\mathbf{g}$ is the gravitational acceleration, $P_\mathrm{t}$ is the total
pressure, $\mathcal{I}$ is the 
Kronecker tensor, $c_\mathrm{h}$ is the
cleaning speed, and $\tau$ is
the damping time scale for divergence cleaning.
Note that the total energy density $e=\rho (\epsilon+v^2/2)+(B^2+\hat{\psi}^2)/(8\pi)$
includes a contribution from $\hat{\psi}$ in 
addition to the internal energy
$\epsilon$, the kinetic energy, and the magnetic
energy. To reduce numerical dissipation
near the grid axis in our simulations,
we have modified the mesh coarsening
algorithm of \citet{mueller_19a} by implementing
a third-order accurate slope-limited prolongation
scheme. Further details on the MHD implementation will
be presented in a code comparison paper
(Varma et al., in preparation). 
The equations for the electron fraction and
mass fractions are the same as in the hydrodynamic
case \citep{mueller_20}.
Neutrinos are treated using the \textsc{FMT} (fast multi-group transport)
scheme of \citet{mueller_15a}, which solves the energy-dependent
zeroth moment equation for three neutrino species
in the stationary approximation using a closure
obtained from a two-stream solution of the Boltzmann equation
(for details such as the neutrino rates, see \citealt{mueller_15a})
The models
are run with a grid resolution of $550\times128\times 256$
zones in radius, latitude, and longitude (cooresponding
to $1.4^\circ$ in angle)  with a non-equidistant
grid in radius out to $10^5\, \mathrm{km}$,
and an exponential
grid in energy space with $21$ zones from $4 \, \mathrm{MeV}$
to $240 \, \mathrm{MeV}$. We use the equation of state of
\citet{lattimer_91} with a bulk incompressibility of $K=220 \,
\mathrm{MeV}$ in the high-density regime, and the same low-density
treatment as in previous \textsc{CoCoNuT-FMT}
simulations \citep[e.g.,][]{mueller_15a}.

\section{Simulation Results}
The maximum, minimum, and average shock radii 
evolve very similarly in the MHD model and the hydro
model up to $250 \, \mathrm{ms}$ after bounce
(Fig.~\ref{fig:evol}a). Small differences arise
because of stochastic variations during 
prompt convection, which affect the entropy profiles
of the PNS mantle and translate into a slightly smaller PNS 
radius and gain radius in the MHD model. These differences are also 
reflected in the critical ratio between the advection time scale 
$\tau_\mathrm{adv}$  and the heating time scale $\tau_\mathrm{heat}$  that 
quantifies the proximity to runaway shock expansion \citep{buras_06b};
$\tau_\mathrm{adv}/\tau_\mathrm{heat}$ is initially smaller
in the MHD model.

Around $250 \, \mathrm{ms}$, however, the critical ratio 
$\tau_\mathrm{adv}/\tau_\mathrm{heat}$ and the shock radius in
the MHD model overtake the hydro model. At $275 \, \mathrm{ms}$ the 
runaway condition $\tau_\mathrm{adv}/\tau_\mathrm{heat}>1$
is met, and steady shock expansion commences with the 
maximum shock radius reaching $1160\, \mathrm{km}$ by the end of the simulation.
The diagnostic explosion energy \citep{buras_06b} has only reached
$2.3 \times 10^{49}\, \mathrm{erg}$ at this stage, but is growing at a rate of 
$4 \times 10^{50}\, \mathrm{erg}\, \mathrm{s}^{-1}$. No explosion develops in the hydro simulation
in agreement with results obtained with more sophisticated
neutrino transport for the same progenitor \citep{summa_18}.

\begin{figure}
    \centering
    \includegraphics[width=\linewidth]{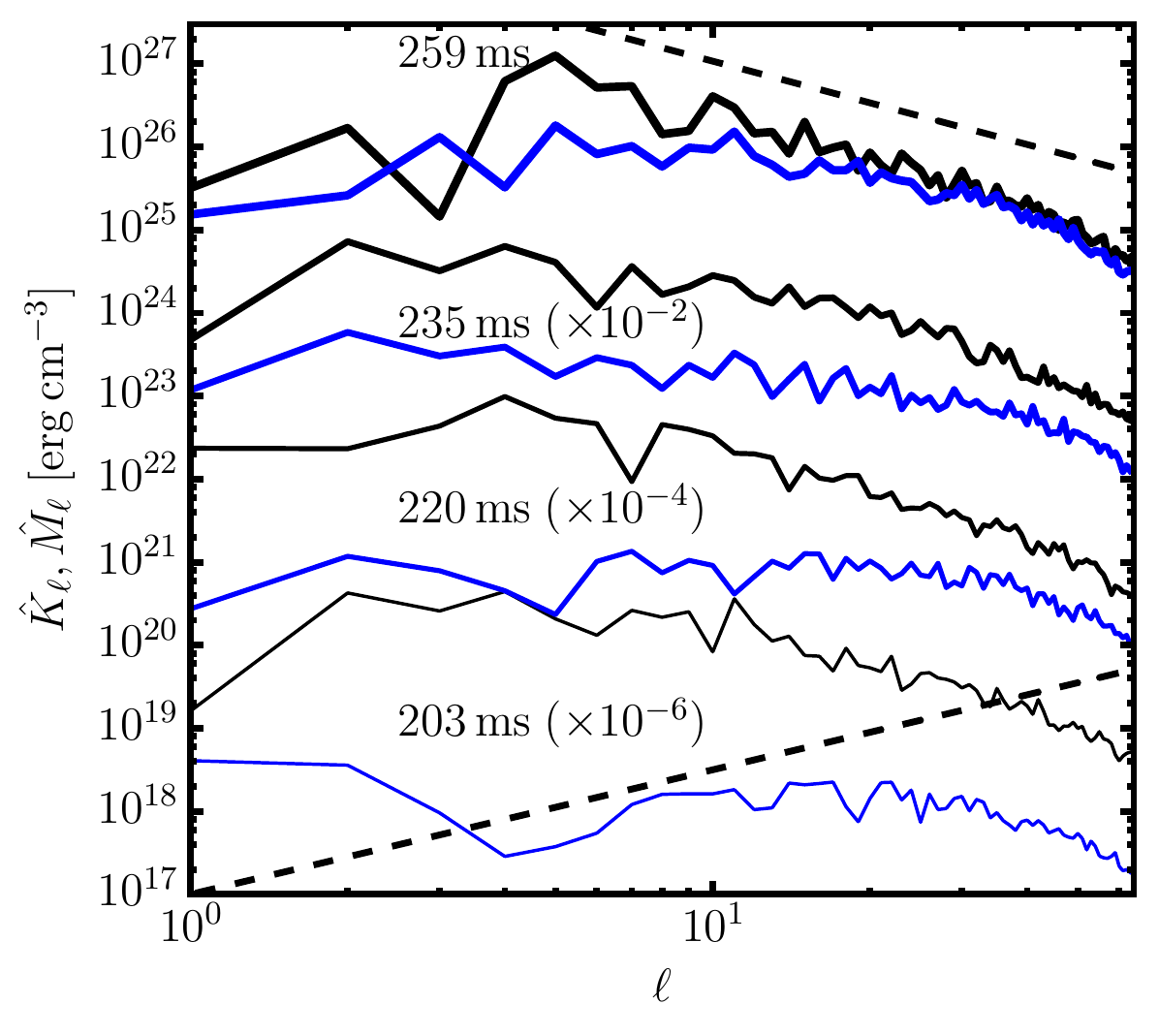}
    \caption{Angular power spectra
    $\hat{K}_\ell$ and $\hat{M}_\ell$
    of the energy contained
    in radial turbulent motions (black) and the
    radial component of the magnetic field (blue) for
    the MHD model at different post-bounce times, measured
    in the lower half of the gain region.
    Dashed lines indicate slopes of $-5/3$ and $3/2$
    for Kolmogorov and Kazantsev spectra.}
    \label{fig:spectra}
\end{figure}

\begin{figure}
    \centering
    \includegraphics[width=\linewidth]{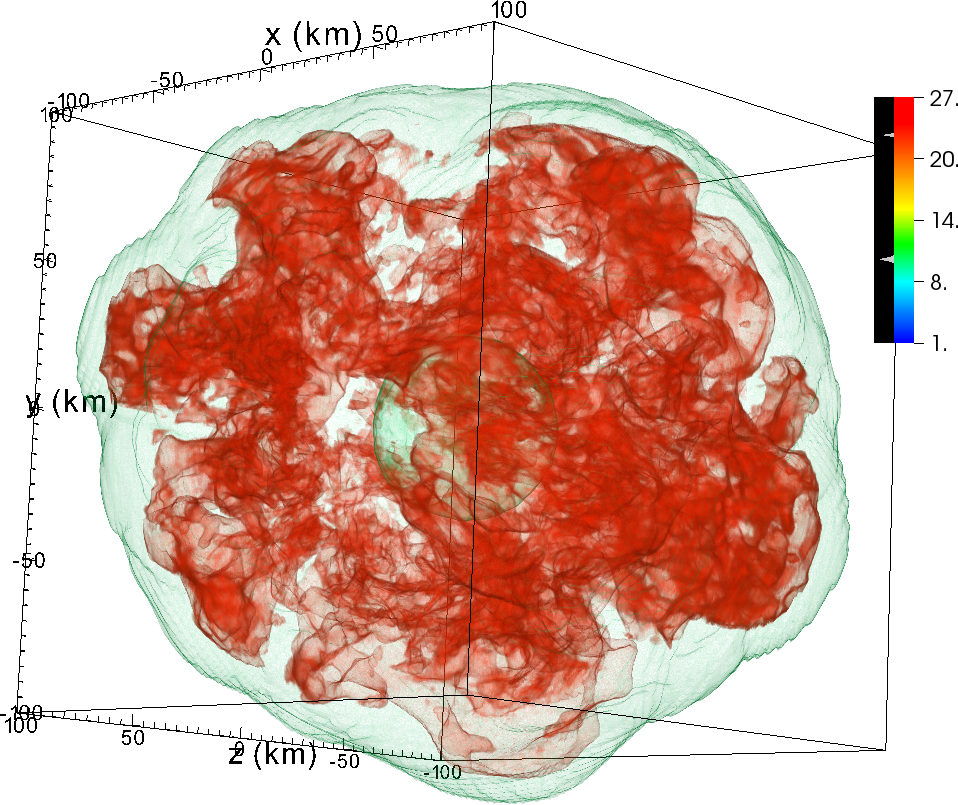}
    \includegraphics[width=\linewidth]{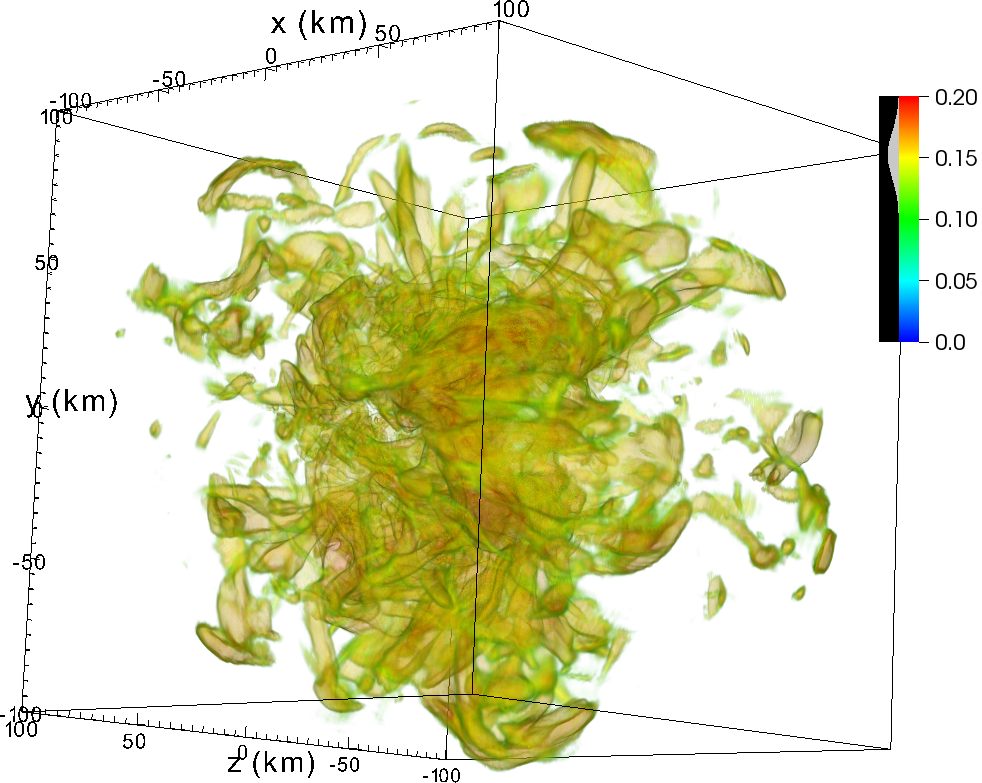}
    \caption{Volume rendering of the entropy
    in  $k_\mathrm{b}/\mathrm{nucleon}$ (top) and the inverse of the plasma-$\beta$ (bottom)
    for the MHD model at a 
time of $260 \, \mathrm{ms}$. Neutrino-heated bubbles are visible
    in red and the shock and PNS surface
    in light green in the top panel.
The magnetic field is dominated
    by small-scale structures. It is distributed
    relatively homogeneously across the gain region, though it tends to be expelled from bigger
    convective bubbles.
    }
    \label{fig:3d}
\end{figure}

The trend towards better heating conditions and shock expansion in the
MHD simulation sets in as soon as the magnetic fields in the gain
region are amplified close to equipartition with the turbulent kinetic
energy. This is illustrated by Fig.~\ref{fig:evol}c, which shows the turbulent kinetic energy
$E_\mathrm{turb}$ in the gain region, and the magnetic
field energy $E_B$ between the shock and the radius where
the cooling rate per unit mass peaks. 
$E_\mathrm{turb}$ and $E_B$ are computed as
\begin{equation}
E_\mathrm{turb}=\int \frac{1}{2}\rho |\mathbf{v}'|^2 \, \ud V, \quad
E_B=\int \frac{|\mathbf{B}|^2}{8\pi}  \,\ud V,
\end{equation}
where $\mathbf{v'}$ denotes the fluctuations of $\mathbf{v}$
around its mass-weighted spherical average.
Exponential field
amplification starts once convection in the gain region develops
and proceeds at a growth rate of the order 
of the inverse convective turnover time, which conforms to the
expected behavior of a small-scale dynamo 
\citep{schober_12}. The growth rate initially increases with time
as the shock contracts and convection becomes increasingly violent.
About $210 \, \mathrm{ms}$, the growth of $E_B$ slows down and
is roughly linear as expected for a small-scale dynamo
after kinetic equipartition is reached for the highest wave numbers
\citep{cho_09}. The ratio $E_B/E_\mathrm{turb}$ peaks
at  $50\%$ around shock revival, i.e., at about the
expected level in the relevant regime of high magnetic Prandtl 
number \citep{schober_15} that is relevant in the supernova
core. The kinetic and magnetic energy spectra also exhibit
the characteristic features of small-scale dynamo amplification.
Fig.~\ref{fig:spectra} shows angular power spectra 
$\hat{K}_\ell$ and $\hat{M}_\ell$ of the kinetic
and magnetic energy contained in the radial components
of the velocity field and magnetic field,
\begin{eqnarray}
\label{eq:kell}
\hat{K}_\ell &=&
\frac{1}{2}
\sum_{m=-\ell}^\ell \left|\int Y_{\ell m}^* (\theta,\varphi) \sqrt{\rho} v_r \,\ud \Omega\right|^2,\\
\label{eq:mell}
\hat{M}_\ell &=&
\frac{1}{8\pi}
\sum_{m=-\ell}^\ell \left|\int Y_{\ell m}^*(\theta,\varphi) B_r \,\ud \Omega\right|^2,
\end{eqnarray}
which we compute in the middle of the lower half of the gain region. 
During the exponential growth phase
there is (leaving aside some jitter at low $\ell$)
initially a peak at high $\ell$ in $\hat{M}_\ell$ and a slightly positive
spectral slope below, which is a little flatter than
the expected Kazantsev spectrum with a power-law slope
of $3/2$ \citep{brandenburg_01,brandenburg_05}, though the small spectral
range does not permit a precise determination of the power-law slope.
Once  $\hat{M}_\ell$ approaches $\hat{K}_\ell$ at high wave numbers,
the growth rate of the turbulent magnetic energy slows down
and a Kolmogorov-like spectrum emerges, though $\hat{M}_\ell$ retains a shallower slope than $\hat{K}_\ell$.
While the dipole component is
non-negligible, the spatial configuration of the field remains dominated by
small-scale structures with little global anisotropy (Fig.~\ref{fig:3d})

Although the behaviour of the MHD model is compatible with
the gross features of a small-scale dynamo, there are
subtle differences to field amplification by 
isotropic and homogeneous turbulence. Amplification is driven mostly by shear motions around the gain radius, and 
the lowest values of the plasma-$\beta$ are reached in this layer
in line with the notion of 
 flux expulsion
from convective regions \citep{weiss_66}. However, the fields
approach or even exceed equipartition with the thermal energy in
strongly magnetized filaments
(Fig.~\ref{fig:3d}) whose volume fraction increases
with time.

It remains to be discussed why magnetic field amplification close
to kinetic equipartition results in more favorable conditions
for shock revival. Supernova theory has established that
the beneficial role of \emph{hydrodynamic} ``turbulence''
(in the broad sense of deviations of the flow from 
spherical symmetry) is through Reynolds stresses and turbulent
heat transfer between the gain radius and the shock,
and that the size of these beneficial effects depends on
the turbulent kinetic energy \citep{mueller_15a,mueller_20,couch_14}.
Similarly, beneficial effects of magnetic fields, e.g., an extra
contribution of magnetic pressure and a reduction of
the binding energy of the gain region, should scale with
the magnetic energy. Although it is by no means clear that adding
the same turbulent kinetic or magnetic energy has the
same impact on the explosion conditions, it is therefore
instructive to compare the total turbulent kinetic and magnetic
energy in and immediately below the gain region in the MHD model
and the hydro model. Since the turbulent motions are driven by neutrino
heating, we consider the dimensionless efficiency
parameter
$\eta_\mathrm{conv}=E_\mathrm{turb}/(\dot{Q}_\nu \Delta R)^{2/3}$ \citep{mueller_15a},
where $\dot{Q}_\nu$ is the volume-integrated heating rate in the gain
region and $\Delta R$ is the
width of the gain region.  Fig.~\ref{fig:evol}d shows that, when taking only the turbulent
kinetic energy into account, $\eta_\mathrm{conv}$ is
very similar in the magnetic and non-magnetic case, reaching a plateau
at $\eta_\mathrm{conv}\approx 0.5$ once convection has fully developed.
If the magnetic energy is included,
$\eta_\mathrm{conv}=(E_\mathrm{turb}+E_B)/(\dot{Q}_\nu \Delta R)^{2/3}$
reaches significantly higher values. In other words, a larger amount
of total turbulent energy can be stored in the gain region for the same
neutrino heating rate if magnetic fields are present; it is not that
the same amount of energy needs to be shared between turbulent 
motions and the additional degrees of freedom in the system. This is in 
line with the finding that the magnetic contributions to the turbulent
energy flux in the gain region almost vanishes. Maintaining balance
between neutrino heating and the turbulent energy flux in a quasi-steady state therefore requires similar convective
velocities in the magnetic and non-magnetic case, and the turbulent 
magnetic energy will be added on top on a level set by the balance
between field amplification and the backreaction of the fields on the flow.

\section{Conclusions}
Our simulations suggest that magnetic fields can have a substantial
and beneficial impact on neutrino-driven shock revival
even in slowly rotating progenitors with weak seed fields.
Field amplification by a small-scale  dynamo in the
gain region is sufficient to amplify fields almost to kinetic 
equipartition within the first hundreds of milliseconds after bounce. 
Judging by the turbulent magnetic energy that can be reached, adding
magnetic fields does, however, have a smaller impact than going
from spherically symmetric models to multi-dimensional hydrodynamic
models. The effect size of magnetic fields on the explosion conditions
will need to be compared to other factors
that influence the heating conditions, such as general relativity and
the treatment of neutrino transport and neutrino interactions. Considering
their impact on $\tau_\mathrm{adv}/\tau_\mathrm{heat}$, magnetic fields are likely one among many factors 
that can contribute to a similar degree to successful explosions
in generic, slowly rotating supernova progenitors
without qualitatively changing the picture of neutrino-driven explosions.

Our MHD simulation prompts a number of questions for
future research. To avoid an overproduction of magnetars,
one important constraint is that the neutron star magnetic fields created during a typical explosion
must not be \emph{too  strong}. 
In our MHD model, the dipole field strength on a density isosurface
at $\mathord{\sim} 10^{10}\, \mathrm{g}\, \mathrm{cm}^{-3}$ is several
$10^{13}\, \mathrm{G}$ and hence somewhat on the high side for typical
pulsars. However, the final dipole field strength of the neutron star
cannot be confidently predicted from our  short simulation. Turbulent reconnection and field burial \citep{torres_16} may yet bring the dipole
field strength down to lower values. Nor can we exclude that
the  relatively strong dipole has arisen by chance, due to
limitations in numerical resolution, or due to the choice of the initial
field. Clearly, more simulations are needed to determine the robustness
of our results; unfortunately the closest analog to
our models --  the idealized 3D simulations of \citet{endeve_12} -- 
are far too different in design to offer a meaningful point of 
comparison.

Further work is also required to investigate whether our results
are sensitive to  non-ideal effects. Conservative
estimates of the physical viscosity and resistivity
based on \citep{spitzer_65} place the magnetic Prandtl number
in the gain region on the order of
$\mathrm{Pm}> 10^3$. While our ideal MHD simulation comports
with the expecations for $\mathrm{Pm}>1$, the numerical Prandtl number is likely
no more than a few \citep{federrath_11}. The numerical Reynolds
number is also bound to be well below  the physical value of
$\mathrm{Re}\sim 10^{15}$ \citep{abdikamalov_15}. Since the growth
rate of the small-scale dynamo for $\mathrm{Pm}\gg 1$ scales
with $\mathrm{Re}^{1/2}$ during the kinematic phase \citep{schober_12},
saturation on the resistive scale should be
reached almost instantly in nature, but
the question becomes whether the growth
of the field on larger scales during the 
subsequent dynamic phase is slow and whether 
saturation may happen well below kinetic
equiparition under certain conditions 
\citep{schekochihin_02}. Simulations \citep{haugen_04} and
more recent analytic models \citep[e.g.,][]{stepanov_08,schober_15} 
do not support such adverse effects on the growth and 
saturation of the small-scale dynamo for $\mathrm{Pr}>1$.
However, any extrapolation to the physical regime is still far from certain, and substantial neutrino drag \citep{melson_20} in the shear layer at the bottom
of the gain region further complicates the picture. Much remains to be done to
substantiate the interesting prospect that magnetic fields may play a beneficial 
subsidiary role in shock revival next to neutrino heating and 
hydrodynamical turbulence.

\section*{Acknowledgements}
BM was supported by ARC Future Fellowship FT160100035.  This research
was undertaken with the assistance of resources and services from the
National Computational Infrastructure (NCI)
and the Pawsey Supercomputing Centre.

\section*{Data Availability}
The data underlying this article will be shared on reasonable request to the corresponding author.

\bibliographystyle{mnras}
\bibliography{paper}
\label{lastpage}

\end{document}